\documentclass[times,twocolumn,final]{elsarticle}

\usepackage{cag}
\usepackage{framed,multirow}

\usepackage{amssymb}
\usepackage{amsmath}
\usepackage{latexsym}

\usepackage{url}
\usepackage{xcolor}
\definecolor{newcolor}{rgb}{.8,.349,.1}

\usepackage{tikz}
\def\checkmark{\tikz\fill[scale=0.4](0,.35) -- (.25,0) -- (1,.7) -- (.25,.15) -- cycle;}

\usepackage{hyperref}

\usepackage[switch,pagewise]{lineno} 

\journal{Computers \& Graphics}

\begin{document}

\verso{Preprint Submitted for review}

\begin{frontmatter}

\title{Multi-dimensional parameter-space partitioning of spatio-temporal simulation ensembles}%

\author[1]{Marina Evers\corref{cor1}}
\cortext[cor1]{Corresponding author.}
\emailauthor{marina.evers@uni-muenster.de}{M. Evers}
\emailauthor{linsen@uni-muenster.de}{L. Linsen}
    
\author[1]{Lars Linsen}

\address[1]{Westfälische Wilhelms-Universität Münster, Germany}

\received{\today}

\begin{abstract}
Numerical simulations are commonly used to understand the parameter dependence of given spatio-temporal phenomena. Sampling a multi-dimensional parameter space and running the respective simulations leads to an ensemble of a large number of spatio-temporal simulation runs. A main objective for analyzing the ensemble is to partition (or segment) the multi-dimensional parameter space into connected regions of simulation runs with similar behavior. To facilitate such an analysis, we propose a novel visualization method for multi-dimensional parameter-space partitions. Our visualization is based on the concept of a hyper-slicer, which allows for undistorted views of the parameter-space segments’ extent and transitions. For navigation within the parameter space, interactions with a 2D embedding of the parameter-space samples, including their segment memberships,
are supported. Parameter-space partitions are generated in a semi-automatic fashion by analyzing the similarity space of the ensemble’s simulation runs. Clusters of similar simulation runs induce the segments of the parameter-space partition. We link the parameter-space partitioning visualizations to similarity-space visualizations of the ensemble’s simulation runs and embed them into an interactive visual analysis tool that supports the analysis of all facets of the spatio-temporal simulation ensemble targeted at the overarching goal of analyzing the parameter-space partitioning. The partitioning can then be visually analyzed and interactively refined. We compared our approach to alternative methods and evaluated it with experts within case studies from three different domains.
\end{abstract}


\end{frontmatter}


\section{Introduction}

\textit{Simulation ensembles of spatio-temporal phenomena} are common in many areas of science like physics or geosciences and in medicine.
An ensemble consists of several simulation runs where each run depends on different parameters or initial conditions. The main idea behind generating an ensemble is to capture the uncertainty in the choice of correct parameter settings or to capture the dependence of the simulation outcome on the parameters. In the latter case, the main analysis task is to extract this dependence. 

\textit{Ensemble visualization} faces the challenge of connecting the simulation outcome with the parameter space. The parameter space can be understood as a multi-dimensional space where each dimension corresponds to one parameter. It contains a set of multi-dimensional points (the \textit{parameter-space samples}), where each point represents one simulation run. For the presented work, we assumed a multi-dimensional parameter space with a dimensionality that is larger than 3. Due to the complexity and, thus, high computation costs of spatio-temporal simulations, it remains infeasible to sample parameter spaces of very high dimensionality. The domain experts who want to analyze their simulation results want to investigate how the parameters influence the outcome. It is often desired to obtain a segmentation or \textit{partitioning of the parameter space}, where each segment represents a connected region with similar simulation behavior. 
One example would be a blood flow simulation, where the flow behavior may change from laminar to turbulent depending on the parameter choices, i.e., one would have a parameter space that is partitioned into two segments (laminar and turbulent flow). The question would then be, which parameter settings lead to which flow behavior and where is the transition between the two behaviors in parameter space, i.e., one wants to analyze the extent of the parameter-space segments and their interface.

The overarching goal of understanding the partitioning of the parameter space is achieved using three components: First, the similarity between the ensemble's simulation runs is analyzed, leading to a semi-automatic generation of similarity clusters. Second, these similarity clusters induce the parameter-space segments, whose extents and transitions are analyzed using parameter-space visualizations. Third, coordinated interactions of similarity- and parameter-space visualizations allow for an interactive refinement of the partitioning.

The methodological part of the paper starts with a problem statement, see Section~\ref{sec:problem_statement}. In Section~\ref{sec:overview}, we provide an overview of the proposed solution for the stated problem. It includes a novel, \textit{distortion-free visualization} for analyzing the \textit{parameter-space partitions} based on the concept of a hyper-slicer. This view is complemented with a 2D embedding of the parameter-space samples to ease the navigation within the parameter space, see Section~\ref{sec:parameter_space_vis}. For the generation of the parameter-space partitioning, the similarities of the ensemble's simulation runs are analyzed using \textit{similarity-space visualizations} of automatic clustering outcomes in conjunction with spatial visualizations as well as visualizations of the runs' temporal evolution, see Section~\ref{sec:system}. Coupling similarity- and parameter-space analyses in coordinated views allows for comprehension and interactive refinement of the parameter-space partitioning. 

We evaluate our approach on synthetic data and compare it to alternative approaches. We then apply it in three case studies to data sets from different fields. We perform the analytical workflow towards the parameter-space partitioning in joint sessions with domain experts in Section~\ref{sec:case_studies}. 
Our main contributions can be summarized by:
\begin{itemize}
\item We propose a novel visualization approach to \textit{investigate partitions of multi-dimensional parameter spaces} of simulation ensembles, where partitions are imposed by clusters of simulation runs with similar behavior.
\item We link the \emph{parameter-space partitioning} visualization to \emph{similarity-space} visualizations of the spatio-temporal simulation runs and embed them in an interactive visual analysis tool that supports the analysis of all facets of the simulation ensemble.
\item We \emph{compare} our approach to parallel coordinates and scatterplot matrices on  synthetic data. We further discuss the usefulness and intuitiveness of our approach with domain experts from different fields within three real-world application scenarios (blood flow, semiconductor quantum wire, microswimmers).
\end{itemize}

\section{Related Work}
Analysis approaches for \textbf{spatio-temporal simulation ensembles} commonly focus on statistical properties like the mean of an ensemble \cite{potter2009ensemble, sanyal2010noodles}. However, such approaches do neither support a comparative analysis nor allow for determining the parameters' influence.
There is also a wide range of clustering approaches for simulation data available \cite{pretorius2011visualization, Hao2016, Ferstl2017, Ma2019, Kappe2019, Kappe2019a}. However, these approaches do not support a direct analysis of the parameter space.
The approach proposed by Phadke et al.~\cite{phadke2012exploring} includes techniques for comparisons but it is limited to a small number of ensemble members. Eichner et al.~\cite{eichner2020making} present a tool for analyzing ensembles of timelines and investigating the parameter space for segmentation algorithms. However, as this approach uses correlations between timelines, it cannot be transferred easily to spatio-temporal simulation data. Wang et al.~\cite{Wang2019} presented a comprehensive survey about the different aspects of ensemble visualization. Another approach~\cite{kumpf2021visual} analyzes the value distributions in multi-field simulation ensembles but does not consider the dependency on input parameters.
A comparative visualization of the whole ensemble is possible with multi-run plots as proposed by Fofonov et al. \cite{fofonov2016visual,fofonovCGF2018}. This analysis approach allows for comparing individual ensemble members, whose temporal evolution is visually represented as a curve in similarity space. We build upon their approach by adopting their field similarity measure and extending it to spatio-temporal similarities.

In ensemble analysis, \textbf{parameter-space analysis} is a key task. Sedlmair et al. \cite{sedlmair2014visual} provided a conceptual framework together with a literature review. Our system mainly follows the navigation strategy global-to-local even though it is common to switch back from a local level to a more global view during the analysis process. The framework includes six analysis tasks of which we focus on the \textit{partitioning task} for which we propose new visualization approaches, but outlier detection via similarity visualizations is also supported.
Clustering ensemble data is a preliminary step to the analysis of the parameter space partitioning. This problem has been addressed by Jarema et al.~\cite{Jarema2015}, who create a distance matrix and use it for hierarchical clustering. However, they neither deal with parameter space analysis nor with parameter space visualization.
Recently, it has been proposed to use machine learning to fill gaps in the parameter space and find interesting regions \cite{Hazarika2020, He2019}. However, these approaches do not include a direct visualization of the parameter space and, thus, do not allow the user to obtain a global overview.

Several approaches for \textbf{parameter optimization} have been proposed. Tuner \cite{Torsney-Weir2011} is a visualization tool based on hyper-slices to investigate the parameter space for image segmentation. They use hyper-slices to evaluate different measures for the quality of the segmentation. In contrast to our approach, they visualize the parameter space concerning quality measures while we investigate a \textit{parameter space partitioning} based on \textit{similarities in spatio-temporal simulation outcomes}. Unger et al.~\cite{unger2012visual} also facilitate parameter optimization by showing the goodness of fit for geoscientific data. For many applications of spatio-temporal simulation ensembles, it is not clear what result is expected and what would be a good measure to quantify how good the results are. Thus, this tool cannot be applied to the problem of generally understanding the parameter space.
Bruckner and Möller \cite{bruckner2010result} propose a tool to facilitate the goal-driven parameter choice for physical animations. They use spatio-temporal clustering to explore the parameter space for the creation of visual effects. However, they do not analyze the relationship between input parameters and simulation output.
Pretorius et al.~\cite{pretorius2011visualization} address the optimization of parameters via a gallery-based visualization for determining parameter settings of image analysis algorithms. However, they only consider a small part of the parameter space because the user's domain knowledge allows limitations. In contrast, we want to analyze the global parameter space to spot interesting features. The local parameter space analysis is also supported by Berger et al.~\cite{berger2011parameterspace}, who use a sensitivity analysis to optimize parameter settings. Even though they provide supporting visualizations of local neighborhoods of points, which should avoid getting lost in the multi-dimensional parameter space, they neither support global overviews nor a geometric understanding of the parameter space.

Obermaier et al. \cite{obermaier2015visual} propose a trend analysis framework and highlight the importance of discovering the dependence on parameters. They also connect their results to the parameter space by using parallel coordinates for \textbf{parameter-space visualization}. Wang et al. \cite{Wang2017} also propose an adapted parallel coordinates plot to visualize the parameter space.
Apart from parallel coordinates, other techniques have been proposed for parameter-space visualization including radial layouts \cite{bruckner2010result} and projections/dimensionality reduction \cite{spence1995visualization,Orban2019}.
 One of our objectives was to obtain an undistorted view of the parameter space, where the values can be read and distances can be interpreted.
Hence, neither parallel coordinates nor dimensionality reduction methods should be used. Glyph-based visualizations are another option\cite{bock2015visual}, but they do not scale to higher dimensions in parameter space.
Instead, we propose to achieve this objective by adopting the concept of \textit{hyper-slices}~\cite{van1993hyperslice}. This concept has been extended to show slices for multiple focus points simultaneously~\cite{Torsney-Weir2018}. However, these visualizations become quickly cluttered and, therefore, are not suitable for our purpose of visualizing partition boundaries. Hyper-slices were also used in HyperMoVal~\cite{Piringer2010} to show the model space. The tool aims to validate regression models and does not include partitioning the parameter space and investigating its properties. 

Other works \textbf{link parameter space and simulation outcome} via iterative selection and refinement of parameter values \cite{Splechtna2015, matkovic2008interactive}. In contrast to our work, they do not aim at getting an overview of the parameter space.
Luboschik et al.~\cite{Luboschik2014} link input parameters to simulation outcome by color-coding the parameters for each ensemble member, which does not provide a geometric overview of the parameter space.
In Paraglide \cite{bergner2013paraglide},
the focus was on sampling aspects and use case evaluations. 
They include a manual parameter space partitioning, while we include a semi-automatic partitioning approach that facilitates working with several similarity clusterings.
Paraglide visualizes the parameter space partitioning using a scatterplot matrix (SPLOM). 
This approach is quite similar to ours, but a SPLOM comes with the issue of overplotting for complete partitions.
We compare our approach against visualizations with parallel coordinates and SPLOMs.

\section{Problem Specification}
\label{sec:problem_statement}
\label{sec:requirement_analysis}

\begin{table}
    \centering
    \begin{tabular}{l|c|c|c|c}
         & \textbf{T1} & \textbf{T2} & \textbf{T3} & \textbf{T4} \\ \hline
       Piringer et al.~\cite{Piringer2010}  &  &  &  & \checkmark \\ \hline
       Splechtna et al.~\cite{Splechtna2015}  &  &  &  & \checkmark \\ \hline
       Matkovic et al.~\cite{matkovic2008interactive}  &  &  &  & \checkmark \\ \hline
       Fofonov et al.~\cite{fofonovCGF2018}  &  &  &  & \checkmark \\ \hline
       Torsney-Weir et al.~\cite{Torsney-Weir2011}  &  &  &  & \checkmark \\ \hline
       Unger et al.~\cite{unger2012visual}  &  &  &  & \checkmark \\ \hline
       Berger et al.~\cite{berger2011parameterspace}  &  &  &  & \checkmark \\ \hline
       Obermaier et al.~\cite{obermaier2015visual}  & \checkmark & \checkmark &  & \checkmark \\ \hline
       Wang et al.~\cite{Wang2017}  & \checkmark & \checkmark &  & \checkmark \\ \hline
       Luboschik et al.~\cite{Luboschik2014}  &  &  &  & \checkmark \\ \hline
       Bergner et al.~\cite{bergner2013paraglide}  & (\checkmark) & \checkmark & \checkmark & \checkmark \\ \hline
       Our approach & \checkmark & \checkmark & \checkmark & \checkmark 
    \end{tabular}
    \caption{Overview of which task is supported by which parameter-space analysis approach. Bergner et al.~\cite{bergner2013paraglide} support partitioning (T1), but only allow for manual partitioning in two groups.}
    \label{table:lit}
\end{table}

\begin{figure*}
    \centering
    \includegraphics[width=\linewidth]{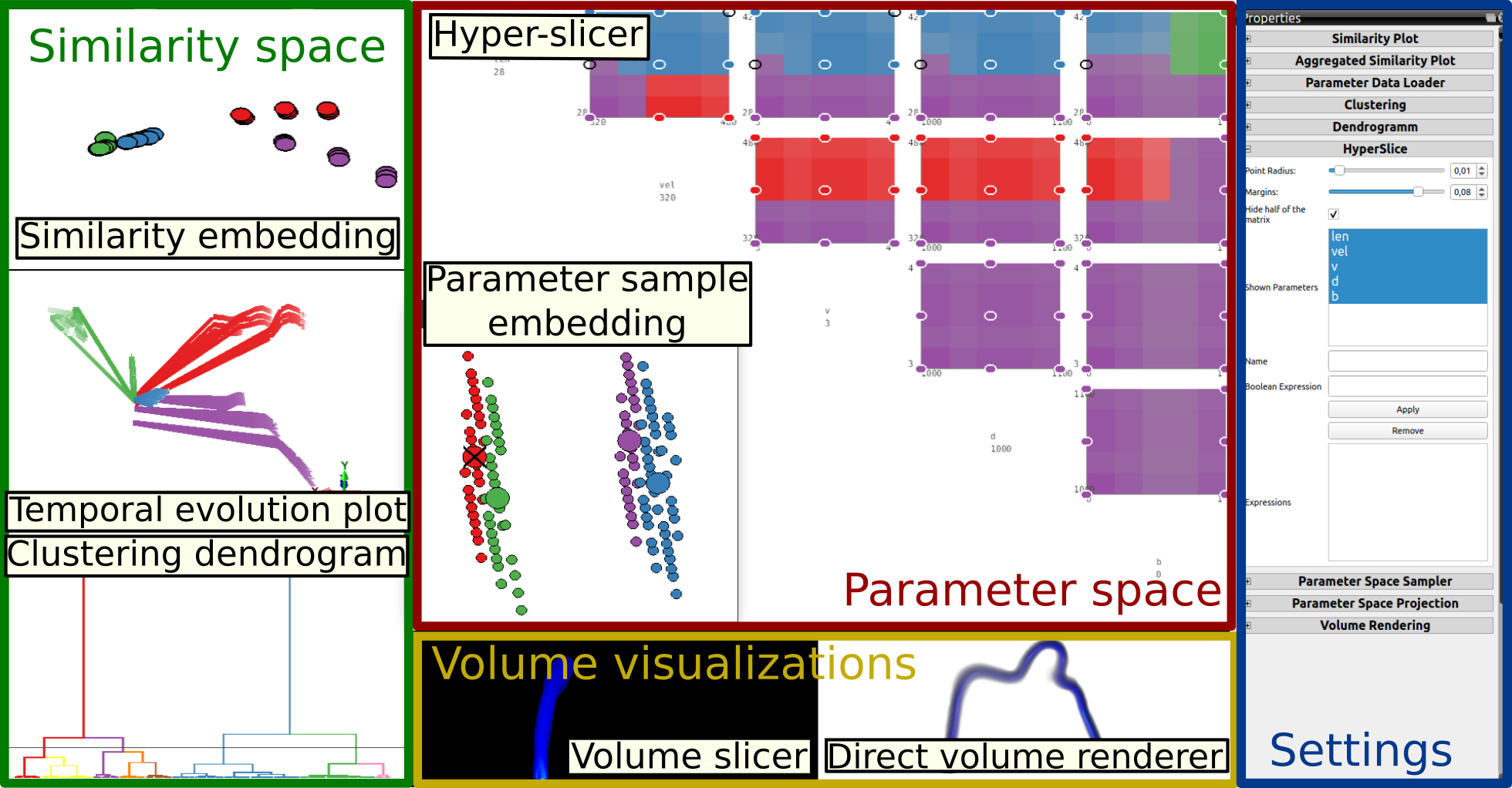}
    \caption{Screenshot of the integrated interactive visual analysis tool. The visualizations show all facets of the simulation ensemble including multiple coordinated views for \emph{similarity space} and \emph{parameter space}, where color is used to visually link the views.}
    \label{fig:overview}
\end{figure*}

We consider an ensemble $E$ consisting of several time-dependent simulation runs $R$, which are also referred to as ensemble members. Each simulation run $R_i$ was created by a simulation with a unique set of input parameter values $P_i$.  The input parameters form the parameter space $P$. In the parameter space, each simulation run corresponds to a single point with coordinates that reflect the chosen parameter values. In the context of this work, we only consider numerical or binary input parameters. The number of input parameters, i.e., the dimensionality of the parameter space $|P|$, may be larger than three, but we assume it not to become too large. Further, each run $R_i$ includes time-varying simulation data, where each time step corresponds to a 2D or 3D scalar field.

The visualization approach should target the following analysis tasks: \\
\textbf{(T1)} \textit{Partitioning} the parameter space based on the simulation outcome: The partitions in parameter space shall be formed by clusters of similar simulation runs. The analysis approach should support the user in defining these clusters and thus creating the partitioning. \\
\textbf{(T2)} \textit{Obtaining an overview} about the parameter space and its partitioning: The parameter space samples should be shown all at once to provide an overview, and it should become clear to which segment of the partitioning the samples belong.\\
\textbf{(T3)} \textit{Analyzing} the extent and the uncertainty of  the different parameter space partitions: The user should understand the geometrical structure of the partitions in parameter space, including their distances, sizes, and shapes. The uncertainties of the partitioning shall be conveyed to avoid misinterpretations. \\
\textbf{(T4)} \textit{Exploring} the simulation outcome on different levels of detail: For a complete understanding of the simulation ensembles and the meaning of the segments, an analysis of the simulation data is inevitable. Here, the analysis of the temporal behavior is important as well as the spatial analysis of individual time steps.

A commonly applied approach to solve these tasks involves a manual analysis by observing each ensemble member individually and comparing the observations. However, this is a tedious and time-consuming process. Thus, the goal is to develop an interactive approach for the visual analysis of the parameter space that incorporates (semi-)automatic algorithms to define and analyze the parameter space partitioning. All methods should be embedded into an \textit{integrated tool} that supports the entire analytical workflow. Table~\ref{table:lit} summarizes, which tools support which tasks. It documents that our integrated tool is the only one that addresses all tasks.
 
\section{Overview}
\label{sec:overview}
To address the analysis tasks listed above, multiple facets of the data need to be investigated, which we propose to do using linked coordinated views of these facets as detailed below. For the analytical workflow, we follow an ``Overview first, zoom and filter, then details on demand"-approach \cite{shneiderman1996eyes}. To have a flexible and extendable tool, we decided to integrate our visualizations into the modular structure of the Voreen framework \cite{meyer2009voreen}.

To support the partitioning of the parameter space (T1), we need to establish a clustering of the ensemble menbers based on the similarity of the simulation outcomes. While cluster generation is a widely used pattern-recognition task and can be performed automatically, any clustering algorithm comes with choices for its settings that influence the clustering outcome~\cite{kumpf2018cluster}. On the other hand, a fully manual cluster generation would require us to inspect all simulation runs in sufficient detail. Hence, we allow for an automatic \textit{cluster generation coupled with interactive analysis and adjustment} of the clustering outcome. We chose a \textit{hierarchical clustering} approach because it only depends on a pruning level in a cluster tree. We support the interactive selection of the pruning level by visualizing the cluster tree in an interactive \textsc{clustering dendrogram}, see Figure~\ref{fig:overview} and Section~\ref{sec:clustering}. Other clustering methods depend on  much less intuitive parameters like the number of expected clusters, kernel sizes, or bin sizes leading to a static output.

To allow for an effective judgment of the clustering outcome, we support its visual inspection. The \textit{similarity space} formed by pairwise similarities of all ensemble runs can be visualized by embedding them into a 2D visual space where each point in the embedding represents an ensemble member. Since we are interested in observing similarities (or dissimilarities), those should be represented as distances in the embedding. Representing dissimilarities as distances in an embedding is obtained by minimizing the objective function of multi-dimensional scaling approach~\cite{wickelmaier2003introduction}. We refer to the visualization of the ensemble members' similarities in an MDS embedding as \textsc{similarity embedding}, see Figure~\ref{fig:overview} and Section~\ref{sec:visualEncodings}. We visually encode the clusters in the embedding by color-coding the sample points.

For an overview of the complete, multi-dimensional parameter space and its partitioning (T2), we propose to use a dimensionality reduction of color-coded parameter space samples $P_i$. We refer to the respective visualization as  \textsc{parameter sample embedding} (see Figure~\ref{fig:overview} and Section~\ref{sec:parameterSampleEmbedding}). For the embedding, we chose MDS because it minimizes stress such that distances are maximally preserved. This view also allows for easy navigation by selecting single samples for a more detailed analysis.

For analyzing the geometric structure of single segments (T3), a distortion-free visualization of the parameter space is needed. This is generally not fulfilled by embeddings that do not allow for reading off the dimensions' values. Parallel coordinates or SPLOMs are alternatives that support this task. However, they suffer from overplotting  and, thus, hinder an intuitive understanding of the geometry of the underlying space as shown in Section~\ref{sec:comparison}. Therefore, we propose a \textsc{hyper-slicer}, see Figure~\ref{fig:overview} and Section~\ref{sec:hyperslicer}.
Hyper-slices provide distortion-free visualizations, which facilitate the required analysis, cf.~\cite{Torsney-Weir2011, Piringer2010}. Additionally, they are based on viewing slices, which is a visual encoding that most domain experts are familiar with because of commonly used slice-based volume viewers. However, existing hyper-slice visualizations only provide information about the selected focus point. We propose a \textit{hyper-slice visualization} that we enhance with information about the \textit{parameter space partitioning} outside of the slice. One remaining issue may be that hyper-slices do not scale well with the number of dimensions. 
We alleviate this issue by including an automatic reordering based on the correlation between the single parameters and the simulation result such that one can concentrate on the most relevant dimensions. 

To understand the simulation behavior over time (T4), we visualize the temporal evolution in similarity space. While the similarity embedding aggregates information over space and time, we also add a visualization that encodes the time component explicitly. Using the same considerations as for the similarity embedding, we create a 2D embedding where the temporal evolution is shown as curves parametrized over time. We refer to the respective view as the \textsc{temporal evolution plot}, see Figure~\ref{fig:overview} and Section~\ref{sec:visualEncodings}. To investigate single time steps, we use \textsc{direct volume rendering} which has become the standard for visualizing volumetric scalar fields. In case of occlusion issues or for 2D simulations, a slice-based visualization referred to as a \textsc{slice viewer} can be used instead or in addition. 

For establishing correspondences between single views, we use consistent color throughout the different visualizations because color is the most intuitive visual variable for this purpose.

\section{Synthetic Dataset}
\label{sec:synthData}

\begin{figure}
    \centering
    \includegraphics[width=\linewidth]{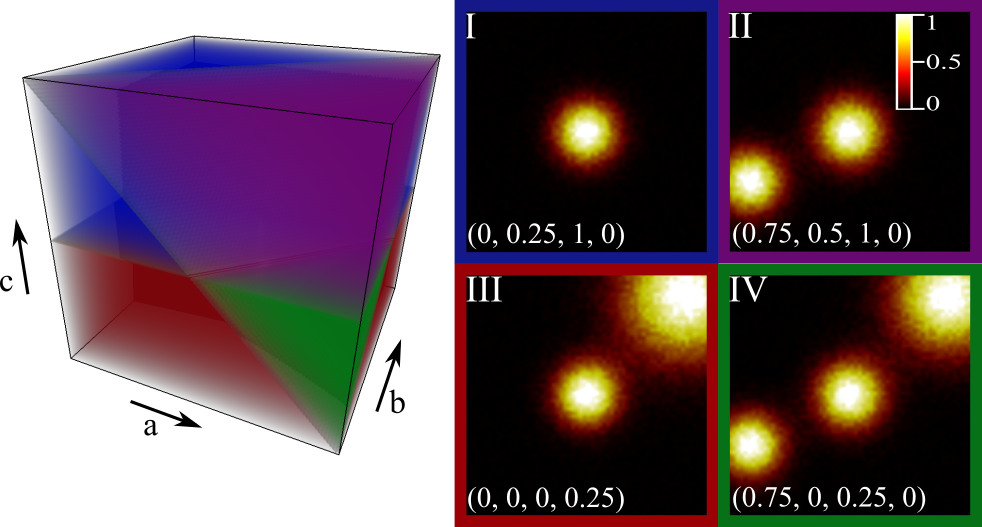}
    \caption{Construction of the synthetic dataset. Parameter-space partitioning with 4 segments (shown in red, green, blue, and purple) created by choices for parameters $a$, $b$, and $c$ (parameter $d$ does not influence the result). I-IV) Example runs for each of the 4 segments, where the corresponding segment is indicated by the colored frames. The exact parameter settings are given in the figure.}
    \label{fig:syntheticdata}
\end{figure}

Throughout the following sections, we will use a synthetic dataset to illustrate the different components, support the explanations of the techniques, and validate our approach. The dataset has a four-dimensional parameter space $[0,1]^4$. Parameters $a$, $b$, and $c$ influence the number and position of Gaussian kernels in a 2D spatial domain (of resolution $64\times 64$), while the parameter $d$ does not influence the output. 
Four different behaviors are modeled, where each behavior is associated with one segment of the parameter space, as shown by the four colors in Figure \ref{fig:syntheticdata}.  We sample the 4D parameter space equidistantly, choosing $625$ samples. The details for the creation of the dataset can be found in Appendix, Section 1.

\section{Parameter Space Visualization}
\label{sec:parameter_space_vis}

Given a parameter-space partitioning (based on a clustering of simulation runs, cf.~Section~\ref{sec:system}), this section is concerned with its visualization.

\subsection{Hyper-slicer}
\label{sec:hyperslicer}
\begin{figure}
    \centering
    \includegraphics[width=\linewidth]{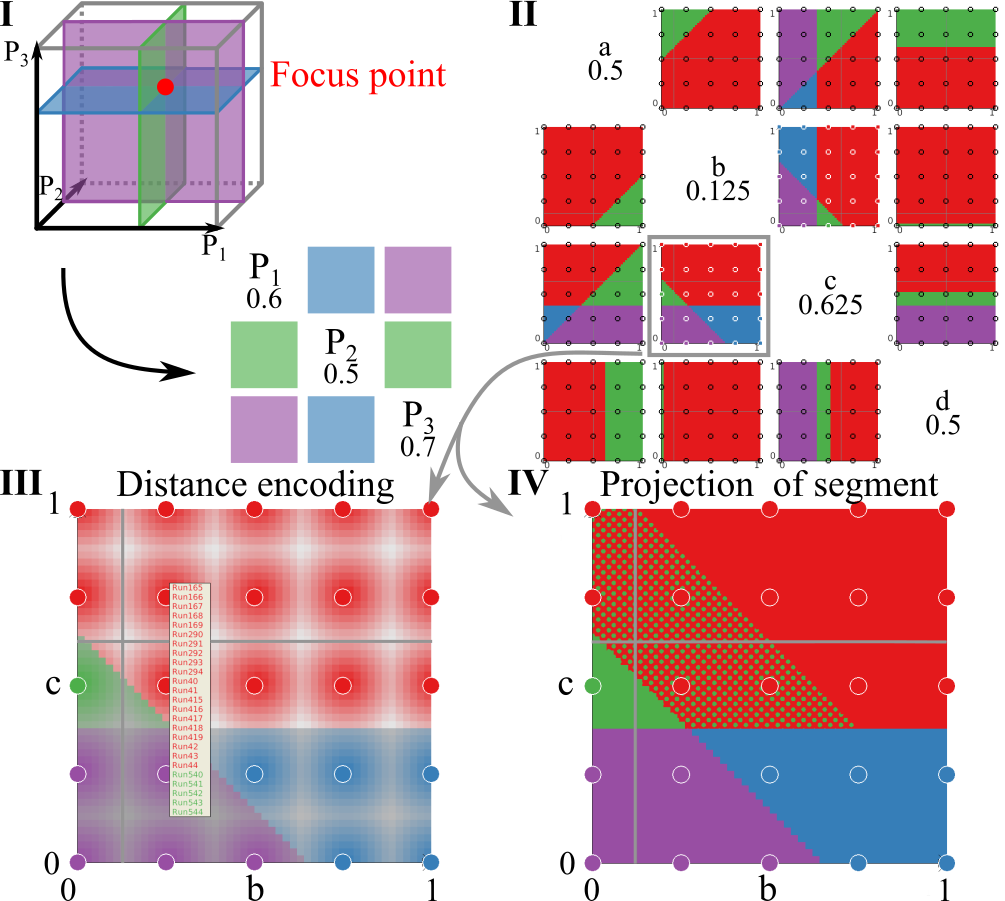}
    \caption{\emph{Hyper-slicer} for visualizing parameter-space partitioning. I) Schematic representation of hyper-slice creation around a focus point (red). The parameters $P_1$, $P_2$ and $P_3$ span a volume (grey cube) from which the slices are extracted. The parameter values together with the coordinates of the focus point are shown on the diagonal. II) Hyper-slicer for the synthetic dataset presented in Section~\ref{sec:synthData}. Sample points inside the slice (white circles) and outside the slice (black circles) are shown. III)-IV) One slice of hyper-slicer with different visual clues: III) focus point (grey lines) and partitioning (colored regions), enriched with uncertainty visualization (saturation) and point labels, IV) enriched with segment boundaries of the selected cluster (green dot texture).}
    \label{fig:slice}
\end{figure}

The visualization of the parameter-space partitioning with undistorted views, preservation of distances, and the possibility to directly read off the parameter values (task (T3)) adopts and extends the concept of the hyper-slice approach suggested by van Wijk and van Liere \cite{van1993hyperslice}. The main idea of hyper-slices for the visualization of a multi-dimensional space is to show axes-aligned slices through all pairwise combinations of dimensions in a matrix-like layout, i.e., a layout similar to SPLOMs. However, in contrast to SPLOMs, hyper-slices do not show a projection of all data samples, but only the selection that is in the slice. Thus, the hyper-slice approach corresponds to a generalization of a volumetric slice viewer to higher dimensions, where multiple slices are shown simultaneously. Treating the parameter space as a multi-dimensional space where each parameter represents one dimension and corresponds to one axis of the multi-dimensional Cartesian coordinate system of the parameter space, we slice the parameter space around a focus point that can be chosen and changed interactively by the user.
This procedure is shown schematically in Figure~\ref{fig:slice}I for a three-dimensional space. The hyper-slice viewer then shows all axes-aligned 2D planes that contain the selected focus point. In each slice viewer, two parameter values are varied on the axis while the others are kept fixed. Those fixed values form the focus point, which represents one parameter setting and is shown by the grey cross. The exact parameter settings of the focus point are additionally shown below the parameter names on the diagonal of the matrix layout.

\smallskip
\noindent
\textbf{Partitioning Visualization.} We extend the hyper-slice approach \cite{van1993hyperslice} to enhance the hyper-slices with the partitioning information. We include the samples' memberships to segments, the segments' boundaries, and respective uncertainties. We show our visual encodings in Figure \ref{fig:slice}.

First, all simulation runs whose parameter settings belong to the selected slice are drawn as colored dots with a white outline that separates them from the background, see Figure~\ref{fig:slice}. 
The colors indicate the cluster membership. Each similarity cluster is assigned a unique color and all parameter-space samples whose simulation runs belong to a similarity cluster are assigned the same color.
We draw a black circle at all projected locations of simulation runs that lie outside the selected slice. This provides context about the \textit{locations in parameter space} of other simulation runs and gives a hint to the user, for which of the parameter settings simulation runs exist. As the parameter space is often sampled on an axis-aligned grid, multiple samples may be projected to the same position within a slice (see Figure~\ref{fig:slice}III). We show the names of all runs when hovering over the position, where the names are written in the corresponding cluster color.
In summary, the visual appearance of the \textit{parameter-space samples' visualization} is similar to a SPLOM with the points lying in the currently selected slice being highlighted.

The \emph{parameter-space partitioning} is visualized by color-coding each slice  according to the similarity clusters.
To obtain the segmentation, we use multi-class support vector machines (SVMs) (see Figure~\ref{fig:slice}II) as implemented in \texttt{libsvm} \cite{libsvm}. We train the SVM based on the clustering in the similarity space. For the training, we use radial basis functions as a kernel. The parameters $C$, defining the costs for wrong classifications, and $\gamma$, which can be intuitively interpreted as the range of influence of a single sample, can be defined by the user. However, as we want to avoid wrong classifications, we show a warning if they occur such that the user can adapt the parameters. The use of SVMs leads to significantly smoother boundaries for sparsely sampled spaces when compared to, e.g. an approximation of the Voronoi diagram. In any case, the boundaries that we show are only an approximation.
If the domain experts want to define segment boundaries more precisely, further simulations are unavoidable. Our tool provides valuable hints about where to sample the parameter space for executing further simulation runs and thus refining the segment boundaries.

\begin{figure}
    \centering
    \includegraphics{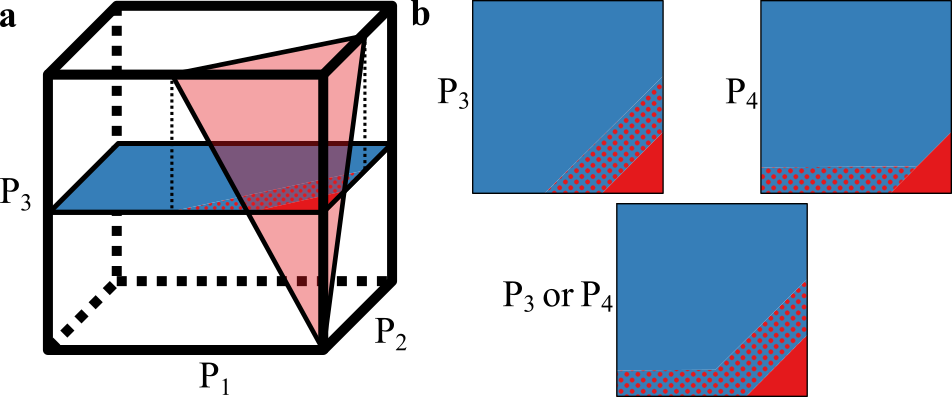}
    \caption{Schematic drawing of boundary projection of red cluster in \emph{hyper-slicer}. a) 3D parameter space with red and blue segments separated by the bright-red triangle. The occurrence of the red segment is projected on the plane indicated by a dotted texture. b) The projections for parameter $P_3$ as in a) and for another parameter $P_4$ as well as the resulting projection for Boolean operation ``$P_3$ \textit{or} $P_4$".}
    \label{fig:schemaProjection}
\end{figure}

We discretize the parameter space to a regular grid for fast computations of the segment boundaries and use the trained SVM to determine to which cluster the grid point belongs. The user can interactively choose the resolution of the grid. A relatively low resolution suffices for most applications, as  parameter-space sampling is often sparse.
Using low resolutions, the visual output of the color-coded slices may exhibit stair-case artifacts (cf.~Figure~\ref{fig:slice}III).
The artifacts can be reduced by using a higher grid resolution, possibly at the expense of a short delay.

\smallskip
\noindent
\textbf{Uncertainty visualization.} Since the parameter-space segmentation partitions the entire parameter space, each grid point in parameter space gets assigned a color no matter whether there was any simulation run located close by. Hence, parts of the parameter space may get colored despite having no simulation run with parameter settings in those parts. To provide a visual hint, we add an \textit{uncertainty visualization} to our slices.
We support the evaluation of the expressiveness by varying the saturation of the colors in the slice viewer (see Figure~\ref{fig:slice}III). %
The saturation is adapted according to the Euclidean distance to the closest parameter-space sample.
We first normalize the parameter values, calculate the distances, and then normalize the distances $d$ to the unit interval.
The saturation of the colors is then decreased by multiplying it with $1-d$. Thus,  lower saturation of the colors corresponds to higher uncertainty. In Figure~\ref{fig:slice}III, which shows a regularly sampled parameter space, one can see that saturation is lowest in between the sample points, i.e., there the uncertainty is highest. The colors are most saturated close to the sample points indicating a low uncertainty. This corresponds to the intuitive expectations that the uncertainty decreases in regions without sample points. The uncertainty visualizations are especially helpful for irregularly sampled parameter spaces, where it is unclear if any sample points outside the selected slice are close to the observed region.
The desaturated colors in uncertain regions remind the user not to over-interpret the clustering results. In fact, segment boundaries become less visible in uncertain regions.
Of course, the uncertainty visualization can be turned off at any time.

\noindent
\textbf{Boundary projection.} A hyper-slice visualization of the partitioning per se only shows the segmentation within the chosen slices. To understand the whole parameter space, one would need to traverse it in all dimensional directions, which can be time-consuming and may impose a high cognitive load. We support this task by providing additional information about the investigated segment within the slice view.
More precisely, we include a \textit{projection of segmentation boundaries} for a selected cluster to guide the exploration of the parameter space. This provides a hint about which parameters need to be varied to observe the shape of the segment. We encode the existence of clusters by using a texture overlay (see Figure~\ref{fig:slice}IV). The texture we chose exhibits dots on a grid rotated by $45^{\circ}$ and colored in the same color as the selected similarity cluster to encode the cluster's extent directly. The dotted texture leads to a perceived extension of the selected segment while keeping the segmentation in the selected slice visible. For the projection of a segment with label $l$ along parameter $P_k$, we create a discrete binary mask $b_{ij}^k$ for each slice that is spanned by parameters $P_i$ and $P_j$, $i,j\in\{1,\ldots,n\}$. Without loss of generality, assume $i<j<k$. Further, let $(f_1,\ldots,f_n)$ be the current focus point. The mask indicates, whether a point $(x,y)$ of the slice shall be textured. For each point $(x,y)$, the mask is computed by

\begin{equation*}
b_{ij}^k(x,y) = \left\{
\begin{array}{ll}
1 & \textrm{if }~\exists z: p(f_1,\ldots,x,\ldots,y,\ldots,z,\ldots,f_n)=l \\
      0 & \textrm{otherwise} \\
\end{array} 
\right.
\end{equation*}

where $p(f_1,\ldots,f_n)$ is the labeled $n$-dimensional parameter space.
The projection itself is represented schematically in Figure~\ref{fig:schemaProjection}(a) for a 3D parameter space (for illustration purposes only). The parameter space is partitioned into two segments by the bright-red plane. Now, the red segment is selected for projection into the slice spanned by parameters $P_1$ and $P_2$ along dimension $P_3$.
The boundaries of the red segment in dimension $P_3$ are extracted and projected onto the slice. The projected area within the boundary is shown with the texture using red dots.
The same result is shown on the left-hand side of Figure~\ref{fig:schemaProjection}(b).
If we have a fourth parameter $P_4$, we could switch to a projecting along that dimension $P_4$ and visualize the respective boundaries of the red cluster. A possible result is shown on the right-hand side of Figure~\ref{fig:schemaProjection}(b). 

Finally, we can also combine projection dimensions using Boolean operations.
If we want to consider the union of the segment boundaries in two  dimensions $P_3$ and $P_4$, we select ``$P_3$ \textit{or} $P_4$", extract the boundaries of the union, and visualize them as shown at the bottom of Figure~\ref{fig:schemaProjection}(b). The union is computed by executing an \textit{or}-operation on the two binary masks.
Any combination of arbitrarily many parameters (apart from the dimensions $P_1$ and $P_2$ that span the plane) can be combined using Boolean operations.
We support the most common Boolean operators including \textit{and}, \textit{or}, \textit{not}, \textit{xor}, \textit{nor}, \textit{nand}, implication, and equivalence.
They are entered within the graphical user interface using a command line.
If the projection for the complete parameter space is desired, the keyword \textit{Complete} can be used in the Boolean expression and the occurrence of segments from all possible parameter combinations is visualized.

We analyze the results for synthetic data to verify our approach. Figure~\ref{fig:slice}II shows the whole hyper-slicer for the entire parameter space and provides an overview. Especially the slice spanned by $b$ and $c$ shows all clusters. We enlarge this slice and activate the uncertainty visualization (see Figure~\ref{fig:slice}III). We can see that the saturation decreases with an increasing distance to the sample points, as expected. Based on the colored labels, we can also already see that runs of the red segment are projected to the segmented sample point as well as runs of the green segments. We use the projection of the segments to see the full extent of the green cluster as shown in Figure~\ref{fig:slice}IV. We see that it is found for larger c values and its boundary is diagonal in this slice. This fits the definition of the dataset.
We also observe that parameter $d$ has no impact. 

\smallskip
\noindent
\textbf{Dimensionality reduction.} An obvious general drawback of the hyper-slice approach is that it does not scale to very high dimensionality. However, in the context of parameter-space visualization for spatio-temporal simulation ensembles, the number of parameters does not become very large. Still, it may be desired to keep the number of dimensions as low as possible. We provide a method to \emph{reduce the dimensionality} of the parameter space interactively by having the user select which parameters to include in the analysis. To facilitate the selection, we propose to re-order the parameters based on the absolute value of the correlation between simulation data and each parameter. This calculation is determined by computing the Pearson correlation coefficient $C_{P_i}$ between the parameter values $P_i$ and the first principle component $v$ taken from the similarity embedding (see below) as 

\begin{equation*}
C_{P_i} = \frac{\sum_{r=1}^n (P_i(r)-\mu_{P_i})  (v(r)-\mu_v)}{\sqrt{\sum_{r=1}^n (P_i(r)-\mu_{P_i})^2}  \sqrt{\sum_{r=1}^n (v(r)-\mu_v)^2}} \ ,
\end{equation*}

where $\mu_{P_i}$ and $\mu_v$ are the mean of $P_i(r)$ and $v(r)$, and $r$ denotes the ensemble member. We visualize the correlations for all dimensions using a bar chart, which supports the user in deciding how many dimensions to include in the analysis, see Figure~\ref{fig:bloodflow} (bottom left).

\begin{figure}
    \centering
    \includegraphics[width=\linewidth]{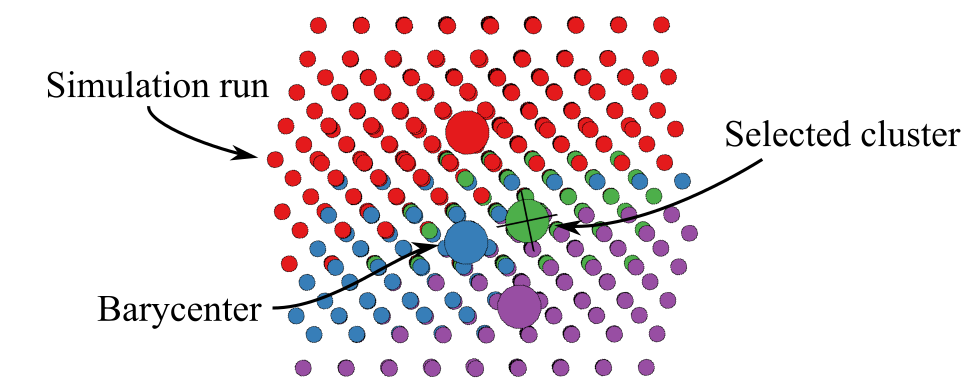}
    \caption{\emph{Parameter-sample embedding} of simulation runs: The four colors indicate cluster membership. Large points represent the clusters' barycenters used for hyper-slice view navigation.}
    \label{fig:mds}
\end{figure}

\subsection{Parameter Sample Embedding}
\label{sec:parameterSampleEmbedding}
To get an overview of the parameter-space samples, we propose to use a 2D embedding of the parameter space based on MDS projections (task (T2)), using the \texttt{R} implementation. This view facilitates navigation in parameter space.
Moreover, it allows for investigating whether similarity clusters also form clusters in parameter space, i.e., it facilitates the evaluation of the quality of the clustering and especially how they are connected in parameter space. 
An annotated example can be found in Figure~\ref{fig:mds}.
Each cluster is represented by drawing the points in the cluster color and by an additional point showing the cluster's barycenter.
The barycenters, shown as larger dots in the cluster color, are added to facilitate selecting the middle of the cluster as the focus point for the hyper-slice visualization.
Since adding the barycenters shall not affect the projection layout, especially not when interactively changing the clusters, the barycenters are computed after projection.

The parameter sample embedding for the synthetic dataset presented in Figure~\ref{fig:mds} shows the regular structure of the parameter space. One can also see that the segments are connected even though there is an overlap. This overlap is expected by the dimensionality reduction but can be resolved by the use of the hyper-slicer.

\section{Ensemble Analysis}
\label{sec:system}
To support the whole analysis process for spatio-temporal simulation ensembles, we embed the proposed parameter-space visualizations described above into an interactive visual system with coordinated views, see Figure~\ref{fig:overview}. The system allows for a detailed analysis of all facets of the ensemble. It also  supports similarity-based clusterings of simulation runs and its interactive adjustment to induce a parameter-space partitioning.

\subsection{Clustering}
\label{sec:clustering}
\noindent
\textbf{Similarity Measure.} Clustering of simulation runs shall be based on their similarity. In order to define and establish a similarity space, we first need to define a similarity measure.
Fofonov and Linsen \cite{fofonovCGF2018} proposed a field similarity measure that determines the similarity between scalar fields as a generalization of an isosurface similarity measure. They presented evidence that its behavior is preferable over other similarity measures in preserving the characteristics of the observed phenomena in ensemble data sets. For a detailed discussion, we refer to their paper. 
A Monte Carlo approach allows for fast computations using random sample points.
From these sample points, vectors that describe the scalar field can be created. The distance $d(V(i, t_m), V(j, t_n))$ between two timesteps $t_m$ and $t_n$ of runs $i$ and $j$ characterized by vectors $V(i, t_m)$ and $V(j, t_n)$ can be calculated as

\begin{equation}
\label{eq:dis}
    d(V(i, t_m), V(j, t_n))=1-\frac{\sum_{k}(1-\max (V_k(i, t_m), V_k(j, t_n)))}{\sum_{k}(1-\min (V_k(i, t_m), V_k(j, t_n)))}.
\end{equation}

\begin{figure}
    \centering
    \includegraphics[width=\linewidth]{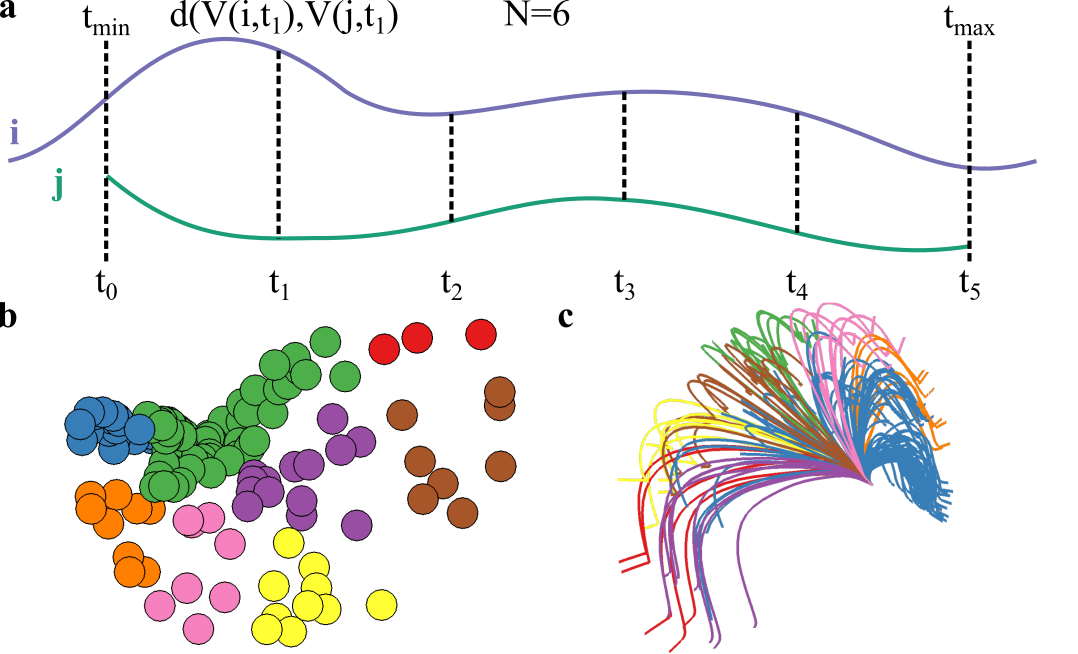}
    \caption{Similarity-space visualizations, where distances encode similarity and colors encode similarity clusters.  a) Schematic representation of the calculation of the aggregated similarity. b)  \emph{Similarity embedding} showing one point per run. A small distance between the points means that the corresponding runs are similar. c) \emph{Temporal evolution plot} showing each simulation run as time curve (here using a 3D embedding). Similar curves represent a similar simulation outcome.
    }
    \label{fig:similarity}
\end{figure}

We generalize this similarity measure for simulation runs.
To establish a fast similarity measure between two runs, we calculate the average distance over time as shown schematically in Figure~\ref{fig:similarity}a. Different runs may cover different time intervals and may have different step lengths. When comparing two runs, we only consider their overlapping time interval $[t_{\min}, t_{\max}]$. This interval is equidistantly re-sampled with sampling length $\Delta t = (t_{\max}-t_{\min})/(N-1)$, where the number of sampling points $N$ is determined as the maximum number of time steps of the two runs within the considered time interval.
We assume that the distances between consecutive time steps are sufficiently small to allow for their linear interpolation.
Hence, the distance between runs $i$ and $j$ is calculated as

\begin{equation}
    d(i, j) = \frac{1}{N}\sum_{n=0}^{N-1} d(V(i, t_n), V(j, t_n))
\label{eq:dis_run}
\end{equation}
with $t_n = t_{\min} + n \Delta t$.

During the interactive analysis, the user may select a temporal region of interest such that the similarity shall be restricted to a sub-interval.
Having pre-computed the similarities for all time samples as in Equation~\ref{eq:dis}, the re-computation of Equation~\ref{eq:dis_run} for the sub-interval can be performed within interactive rates. We also want to point out that, since we are computing similarities pairwise, we can always use the largest temporal overlap of each pair of runs for the similarity computation.

Depending on the application scenario, it may be desirable to compute the similarities as in Equation~\ref{eq:dis_run} or to compute the best match of the time series by allowing a time shift (up to an upper threshold $\tau_{\max}$) for a temporal alignment.
In the latter case, we replace Equation~\ref{eq:dis_run} by

 \begin{equation}
     d(i, j) = \frac{1}{N}\min_{\tau}\left(\sum_{n=0}^N d(V(i, t_n), V(j, t_n+\tau))\right),
 \label{eq:dis_run_shift}
\end{equation}
 where $\tau$ is varied in discrete steps between $-\tau_{\max}$ and $\tau_{\max}$.

Computing pairwise distances $d(i, j)$ of all simulation runs leads to a distance matrix $D_R$, which spans the \emph{similarity space}.
Which similarity measure  to use (Equation~\ref{eq:dis_run} or Equation~\ref{eq:dis_run_shift})  is decided by the domain expert performing the interactive analysis.

\smallskip
\noindent
\textbf{Hierarchical Clustering.} To detect patterns in the similarity space of the simulation runs, we apply a clustering method based on distance matrix $D_{R}$ (task (T1)).
As it is not a priori known how many clusters exist, we propose to use a \textit{hierarchical clustering} approach.
A hierarchical clustering approach generates a cluster tree by iteratively merging clusters in a bottom-up fashion until all clusters are merged into one cluster, which becomes the root of the tree. The number of clusters can then be determined in retrospect by analyzing the cluster tree.

In hierarchical clustering, the results depend on the choice of the \textit{linkage algorithm}. As the optimal choice for the linkage algorithm is data-dependent and thus cannot be generally answered, we implement the possibility for the user to choose from a list of the most common linkage algorithms.
The users may interactively test multiple linkage algorithms and visually compare their outcomes to choose the optimal algorithm for their goal.
The list of available algorithms includes (i) Ward's minimum variance method that minimizes the increase in variance when two clusters merge (called ward.D2 in the following), (ii) an adaption thereof where the input distances are not squared (ward.D), (iii) single linkage which is based on the minimum distance, (iv) complete linkage which is based on the maximum distance between points, (v) average linkage using the unweighted pair-group method for arithmetic averages (UPGMA) which minimizes the average distance between pairs of elements in the union of the clusters, and (vi) average linkage using the weighted pair-group method for arithmetic averages (WPGMA) which is based on the average distance between pairs from both clusters.
For the implementation of the hierarchical clustering algorithms, we used the \texttt{R} package \cite{R}, which we accessed from our C++ code using the \texttt{RInside} package \cite{RInside}.

\begin{figure}
    \centering
    \includegraphics[width=1.0\columnwidth]{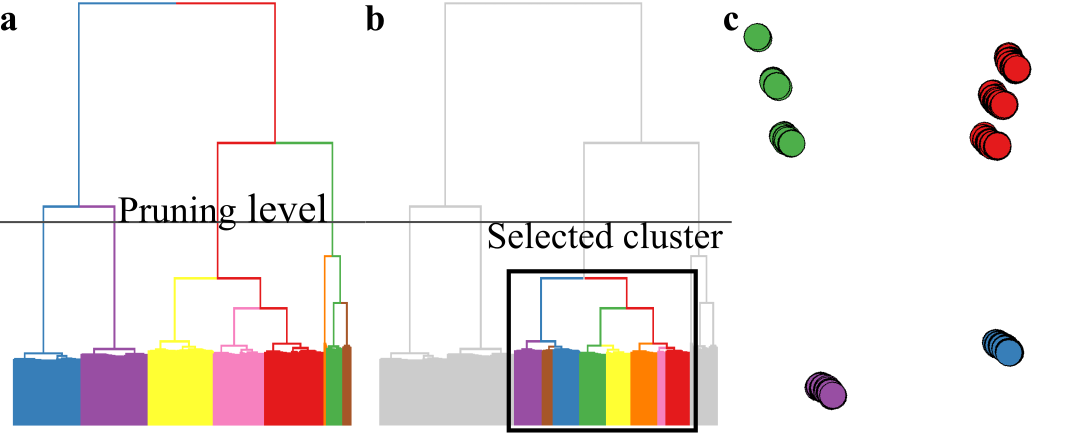}
    \caption{\emph{Clustering dendrogram} for hierarchical clustering. Pruning level is used to interactively decide on the number of clusters. a) Full dendrogram with a color map created using ColorBrewer. b) One cluster selected in dendrogram for further investigation. c) Corresponding similarity-space visualization to the dendrograms.}
    \label{fig:dendrogram}
\end{figure}

\smallskip
\noindent
\textbf{Cluster Analysis.} A hierarchical clustering does not create a single clustering solution. However, it represents a \textit{hierarchical ensemble of clusterings}, from which a single clustering result can be extracted by pruning the cluster tree.
To facilitate the pruning, we visualize the hierarchical clustering outcome with a \textit{dendrogram}. This cluster tree visualization exhibits in which order the clusters merge and split. The height in the dendrogram also conveys the distances at which the clusters change.
The users can prune the cluster tree by interactively adjusting the height of a horizontal line representing the pruning level.
The number of clusters is thus implicitly given by the pruning selection.
For example, in Figure~\ref{fig:dendrogram}a, the number of clusters is four.

We link the interactively chosen clustering result to other views to analyze the clusters and their interactive adjustment further. This is achieved through the use of colors.
We identify three requirements to apply a \textit{color map} to the cluster tree: (1) The colors should be clearly distinguishable. (2) Since clusters correspond to the parameter-space segments, an excessive number of clusters should be avoided, as it would lead to an oversegmentation of the parameter space. (3) When adjusting the pruning level of the cluster tree, the assigned colors should remain unchanged as much as possible.
Hence, we decided to pick the colors from a qualitative (or categorical) color map with up to $12$ clusters generated with ColorBrewer \cite{colorBrewer}. (For the examples shown in this paper, we used the scheme ``Set 1'' with $8$ classes.)
We assign the colors by traversing the cluster tree in a top-down manner. The root node gets assigned the first color. At each inner node, the child with the largest subtree gets assigned the same color as its parent, while the other child gets a new color from the set of still available colors. If no further color is available, both children get assigned the color of the parent.
The procedure progresses until the leaves are reached, see Figure~\ref{fig:dendrogram}a.
It can be observed that clusters at a lower level in the hierarchy will be assigned the same color.
However, it is possible to select sub-clusters for further analysis as in Figure~\ref{fig:dendrogram}b, where colors are re-assigned to the selected sub-clusters, while 
already assigned colors within the selection are maintained.
Not selected sub-clusters are colored in light grey. 

\subsection{Visual Encodings}
\label{sec:visualEncodings}
\textbf{Similarity Embedding.} We introduce a similarity embedding to visualize the \textit{similarity space} of all simulation runs as captured by distance matrix $D_R$.
Hence, we use a low-dimensional embedding, where each point represents a simulation run. The distances in the embedding shall reflect the distances in the distance matrix $D_R$ as much as possible. This goal reflects the objective function of an MDS approach \cite{wickelmaier2003introduction}.
The similarity embedding is linked to the other visualizations via the assigned cluster colors, see Figure~\ref{fig:dendrogram}c. It allows the user to evaluate the quality of the clustering and the user can interact with the dendrogram to adjust the cluster selection for achieving a better matching.
For the synthetic dataset, we observe four separate groups corresponding to the clusters for the pruning level selected in Figure~\ref{fig:dendrogram}a. By choosing a lower pruning level in the dendrogram, we can also confirm that the green and the blue cluster split into three subgroups each.

\smallskip
\noindent
\textbf{Temporal Evolution Plot.}
The clustering approach that induces the parameter-space partitioning is governed by distance matrix $D_{R}$, where we aggregate over time. However, the investigation of the temporal evolution is essential to observe if the runs also behave similarly over time and find out, e.g., if they diverge or converge (task (T4)).

To visualize the temporal evolution, we follow the ideas of the multi-run plot proposed by Fofonov et al.~\cite{fofonov2016visual}.
Therefore, we compute the \mbox{dissimilarities} between all time steps of all runs according to Equation~\ref{eq:dis} and store them in a distance matrix $D_T$.
Hence, we would like to use a low-dimensional embedding, where each timestep of a simulation run is represented by a point and where the distances in the embedding reflect the distances in matrix $D_T$ as much as possible, which is again achieved by an MDS approach \cite{wickelmaier2003introduction}. The projected points are then connected in temporal order to obtain a curve showing the similarity over time and getting an overview of the temporal evolutions of the runs' similarity.
The simulation runs are color-coded according to the clustering, which allows for a quick evaluation of the temporal evolution of cluster members, see Figure~\ref{fig:similarity}c.
For the embeddings, we can use 1D, 2D, or 3D spaces.
Which dimensionality is chosen depends on the intrinsic dimensionality of the data set.
1D embeddings are particularly intuitive, as they can be plotted using time as a second axis. The 1D embedding can also be used to intuitively select a time interval of interest that should be analyzed further. This time interval is shown in 2D and 3D embeddings by rendering the parts of the curve outside the selected time interval with decreased saturation.

\subsection{Interplay of Components \& Analytical Workflow \label{sec:workflow}}

A typical analytical workflow using our tool starts by getting an overview of the ensemble using the \emph{similarity embedding} and the \emph{temporal evolution plot}. The temporal evolution plot can be used to limit the time range taken into consideration for the analysis. Using the \emph{clustering dendrogram} together with the similarity embedding, a suitable linkage algorithm and pruning level of the hierarchical clustering is chosen. This clustering induces a partitioning of the parameter space, which can be observed in the \emph{parameter sample embedding} to get an overview. 
Then, we can investigate this partitioning and the boundaries of the segments in detail and understand their relationship to the parameter values. This is done via the \emph{hyper-slicer}. Using the correlation between parameter values and simulation outcome, we reduce the dimensionality of the hyper-slicer by deselecting uninteresting parameters. Clusters can be selected in the \emph{parameter sample embedding} to change the focus point to the respective cluster in the \emph{hyper-slicer}. The boundaries of this cluster can be investigated to understand its extent in parameter space. For a connection back to the simulation outcome, we select a single ensemble member in the \emph{hyper-slicer} and observe its temporal evolution in a \emph{volume rendering} or a \emph{slice viewer}, which are both common techniques for volume visualization (task (T4)). Using our system, it is easy to change the level of detail of the investigations. Common usage scenarios include, e.g., refining the temporal selection, changing the method or the pruning level of the clustering, or selecting various clusters for in-detail investigations. A detailed walkthrough for a concrete application scenario is presented in Section \ref{sec:bloodflow}.

\section{Case Studies}
\label{sec:case_studies}

\begin{figure*}
    \centering
    \includegraphics[width=0.95\linewidth]{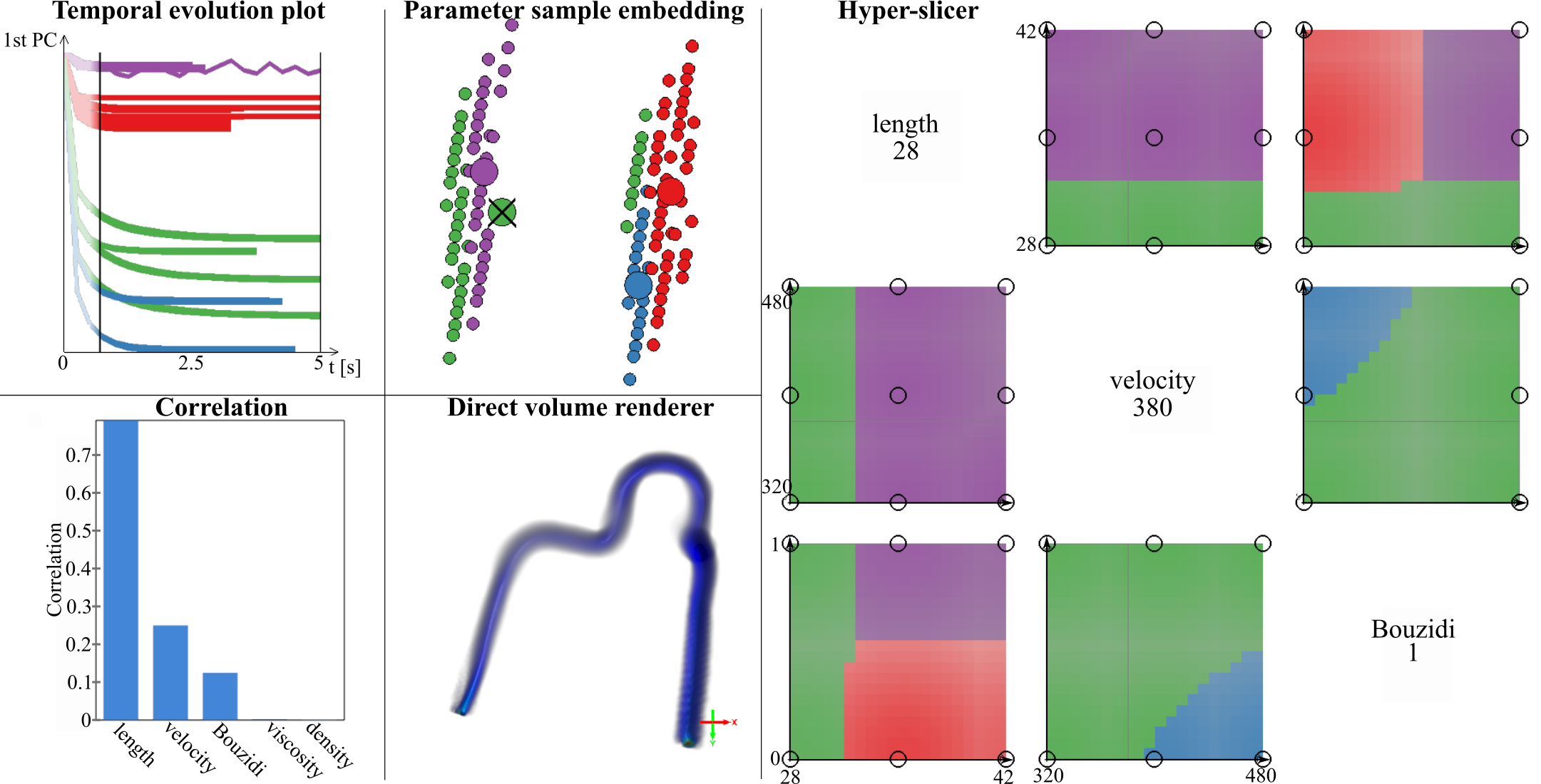}
    \caption{Analysis of the blood-flow dataset for the time interval shown in the \emph{temporal evolution plot}. The ensemble can be divided into four similarity clusters, which in the \emph{parameter space embedding} are mainly separated by the choice of parameters \emph{length} and \emph{Bouzidi}. The respective partitions can be observed in the \emph{hyper-slicer}, for which the number of parameters was reduced based on the \emph{correlation} bar chart between parameters and simulation outcome. Individual time steps of individual runs are investigated using a \emph{direct volume renderer.}
    }
    \label{fig:bloodflow}
\end{figure*}
We present three case studies from different domains, one within this paper and two in the supplementary material.  For the first case study, we include an in-detail walkthrough explaining how the insights were obtained. We also conducted an informal user study with, in total, four domain experts (two professors and two graduate students), which were involved in the creation of the three datasets we used.  We started with a short introduction to our tool and the respective visualizations. Then, they explored their datasets (in the same session) and gave us feedback. 
The timings reported below were taken on a laptop with a 1.6~GHz Intel Core i5 processor.

\subsection{Blood Flow }
\label{sec:bloodflow}

We analyzed the simulation results of blood flow in an aneurysm \cite{leistikow2020interactive}. The scientists want to understand how the parameters influence the simulation outcome to find a suitable match to experimental data. We used the magnitude of the flow field created by using a Lattice Boltzmann method. The simulation outcome depends on five parameters: a characteristic length, a characteristic velocity, the viscosity of the fluid, its density, and a parameter that stores whether Bouzidi boundary conditions, which are a special kind of boundary conditions for fluid simulations, are used. The dataset consists of $129$ runs with $8$ to $42$ equidistant timesteps. Each scalar field has a grid resolution of $128\times128\times128$ sample points. To obtain a sufficiently good representation,  $32,768$ Monte Carlo seed points per timestep are chosen for the similarity calculations.
The relevant time interval for further analysis was determined and interactively chosen to 
only start at simulation time $0.7$~s.
Timings and scaling for computing distance matrix $D_T$ are discussed in detail in literature \cite{fofonov2016visual}. Given $D_T$, the subsequent calculation of distance matrix $D_R$ took $1.0$~s.
We evaluated the application on this dataset with the help of two domain experts, where one of them created the simulation ensemble and the other one ran accompanying measurements.

At first, the similarities were investigated and a suitable clustering based on the similarity plots is chosen. Using our tool, different linkage algorithms were tested and Ward's minimum variance method without squared distances (\textit{ward.D}) was identified to create the most reasonable clusters.
A division into two similarity clusters directly becomes evident from the clustering dendrogram.
(see Figure~\ref{fig:overview}). We selected the partitioning by choosing a suitable pruning level. This clustering was investigated next in the parameter space. Looking at the parameter sample embedding, one can see that the two clusters in parameter space do not agree with those in similarity space. 
Using the hyper-slicer, we observed that the parameter \textit{length} leads to a separation of these two clusters. These findings agree with the expectations of the domain expert. However, he rather had expected the influence of the boundary conditions to be larger than the influence of the length.
Our observation, however, can be confirmed by calculating the absolute value of the correlation between the simulation data and the individual parameters (see Figure~\ref{fig:bloodflow} bottom left). We further observed that the two parameters \textit{density} and \textit{viscosity} play no significant role and decided to exclude them from further analysis.

Next, we intended to refine the two initial clusters. In the clustering dendrogram in Figure~\ref{fig:overview}, we observe that both clusters split into two sub-clusters at approximately the same height (red cluster splits into red and purple, while blue cluster splits into green and blue).
In the parameter sample embedding in Figure~\ref{fig:bloodflow}, we can see that this splitting almost leads to a separation of the two cluster pairs in parameter space. Only the green cluster occurs in both groups. Thus, this green cluster is selected in the parameter sample embedding for further investigation. Projecting the segment to the slice, we observe that no further regions contain the cluster. The hyper-slice view also shows a clear separation of the red and the purple cluster caused by the Bouzidi boundary condition. The green and blue clusters, which occur only for small lengths (i.e., a high spatial resolution of the simulation domain), are not separated by the choice of boundary conditions. Instead, the blue cluster only occurs without Bouzidi boundary conditions and for higher velocities. This observation can be interpreted as a lower influence of the boundary conditions for higher spatial resolutions. To better understand the simulation result, single points in each cluster were selected and investigated using a direct volume renderer.

\subsection{Findings}
The domain experts ranked the tool as very helpful for their work. Our visualizations facilitate the process of understanding the data and the parameter influence. Two of the domain experts explicitly mentioned the importance of linking the abstract visualizations with the spatial visualizations to support an intuitive understanding. The domain expert who ran the blood flow simulations stated that our tool allowed him to confirm prior knowledge like the strong influence of the characteristic length and the boundary conditions. At the same time, he got new insights into the dominance of the individual parameters.
Three domain experts positively highlight the broad applicability of the tool. The one who works with the blood flow dataset as well as with experimental data would like to include her experimental data into the visualization and use the approach to analyze the parameter values of the segment closest to the measurements. The physicist working on microswimmers (see Appendix, Section 3) states that the tool permits new possibilities in analyzing data that depends on more than three parameters. He rates the hyper-slicer as especially helpful to obtain a mental picture of the parameter space's geometry.  It helps him to understand phenomena depending on multi-dimensional parameter spaces, which he usually would limit to fewer parameters. However, the limitation of parameters comes with the risk of missing interesting behavior predicted by the model. The domain scientist working on semiconductor simulations (see Appendix, Section 2) pointed out that the tool saves him time in the analysis process and showed him some parameter dependencies worth a deeper analysis with further simulations. Without our visualizations, he was not aware of this interesting region in the parameter space. He sees an application of our approach in finding parameter sub-spaces for further analysis.

One of the domain scientists would appreciate further visual support in investigating a selected cluster, which we plan to include in future research. Including an uncertainty visualization into the hyper-slicer was also suggested by the user with whom we conducted the first case study. We included this before the sessions with the other experts. The users rated especially the interactive exploration as very helpful and intuitive, and one of them also positively highlighted the high information density. However, it became clear that training is necessary to use our tool. The need for an initial training session is common for most complex tools and does not limit the intuitiveness to a trained user.

\section{\textbf{Comparison to SPLOM and PCP}}
\label{sec:comparison}
\begin{figure*}[!hbt]
    \centering
    \includegraphics[width=\linewidth]{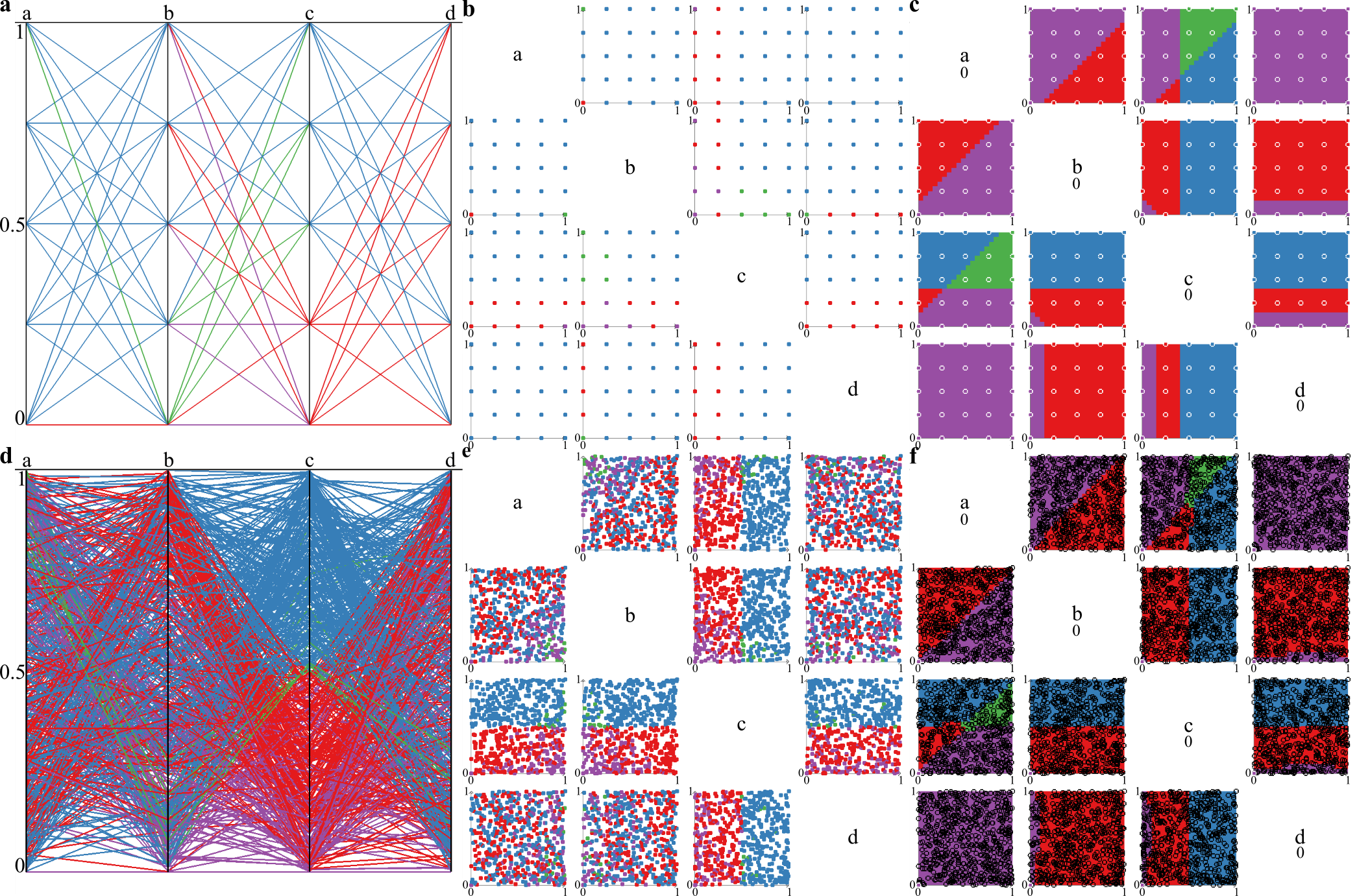}
    \caption{Comparison of alternative visualization approaches for parameter space partitioning. Examples in (a, b, c) show a structured sampling of the parameter space, while those in (d, e, f) show a Monte Carlo sampling. Results for PCP (a, d), SPLOM (b, e), and our hyper-slicer (c, f) are presented. We observe that PCP and SPLOM suffer from overplotting (a, b) and do not exhibit the shape of the segments well (d, e).}
    \label{fig:syntheticResults}
\end{figure*}

We compare our extended hyper-slicer approach to the use of scatter plot matrices (SPLOMs) and parallel coordinate plots (PCPs). Here, PCP and SPLOM are only compared to the visual encoding of our hyperslicer, not as a replacement for our whole  system. Parameter spaces are often sampled on a structured grid, e.g., for the datasets presented in Section~\ref{sec:bloodflow} (Bloodflow) and Appendix, Section 3 (Microswimmers).

The results for a structured sampling of the parameter space when using the three approaches are shown in Figure~\ref{fig:syntheticResults}a-c. As in the hyper-slicer, the segments in the PCP and SPLOM are color-coded. One can directly see that both PCP and SPLOM suffer significantly from overplotting. The structured sampling can explain this. For PCP, it is especially problematic as all possible parameter combinations are present in the dataset. Thus, many lines are drawn on top of each other. This also explains why there are no visible purple lines between $a$ and $b$. Another problem might arise due to misleading information. In both visualizations, it is not directly possible to see how many points or lines are drawn on top of each other. The visualization is also strongly impacted by the rendering order which determines which color is shown. Even though one can suspect from the PCP as well as the SPLOM that the red cluster only occurs for small values of c while the blue one is visible for larger values, it is not clear if this is not a projection artifact due to overplotting. Transparency or order-independent blending might reduce this issue in some cases, mainly if the density is not homogeneous, but it does not generally help in visualizing differently colored clusters and may cause some hard to interpret mixed colors. The hyper-slicer addresses this issue by only showing a clearly determined slice and, thus, a subset of the total information.

In principle, a similar extension of showing subsets can be introduced for PCP and SPLOM as well. However, the hyper-slicer enables the user to develop a geometric understanding. To compare the different approaches for this task, we use an unstructured sampling as presented in Figure~\ref{fig:syntheticResults}d-f. In those cases, overplotting is less prominent for SPLOM and PCP. However, it is still hard to identify some structures. For example, while the separation caused by parameter $c$ can be identified in PCP and SPLOM, the diagonal structure for the green and the purple cluster is hardly visible (see Figure~\ref{fig:syntheticdata} for the shape of the segments in 3D). Different adjustments like edge bundling for PCP or density-based scatterplots for SPLOM might partially improve the perception but will not fully solve the problem. The reason for this lies in the large amount of data which is shown completely in SPLOMs and PCP, while only a subset is visualized in the hyperslicer and this subset can be changed interactively. The local view of the hyper-slicer allows us to spot the diagonal shapes, for example, in the slice spanned by parameters $a$ and $c$. The possibility to interactively change the focus point further supports building a geometric understanding of the parameter space (see video).

\section{Discussion and Conclusion}
\label{sec:discussion}

We presented an approach to analyze simulation ensembles leading to a parameter-space partitioning based on the similarity of the simulation runs. We successfully applied our visualizations to different analysis tasks and the domain experts were able to confirm existing knowledge and get new insights into their data. We received positive feedback together with helpful suggestions. 
It was even mentioned that our approach would allow the analysis of datasets with higher-dimensional parameter spaces, e.g., a ten-dimensional parameter space, which before was mainly avoided due to the high effort of manual analysis that is necessary. Additionally, our visualizations hint at possibly interesting parameter regions that otherwise might be missed in manual analysis.

A well-known problem arising with hyper-slices is the bad scaling to high dimensions known from SPLOMs. Even though this might lead to problems in some specific cases, for most simulation ensembles, the number of simulation parameters is not that high.  For higher-dimensional parameter spaces, a dense sampling is computationally barely possible. Additionally, we included the possibility to determine the correlation coefficient between the single parameter values and the simulation data. This introduces a ranking of the parameters and facilitates the selection of relevant parameters. Thus, the dimensionality of the parameter space can be reduced.
One of our domain experts would have liked to select more than one cluster at a time, which could be easily added to our approach. However, when selecting several clusters simultaneously, the visualizations may quickly become cluttered. Finding a suitable visualization dealing with this problem and facilitating the further analysis of identified clusters will be part of future work.

Our methods are  based on a (dis-)similarity matrix. The choice of a similarity measure is crucial. Depending on the application, there might be cases where the field similarity is not the best choice. To analyze the dependence of the analysis result with our methods on the similarity measure, further research is needed. However, exchanging the similarity measure is straightforward, which then allows for the direct application of our visualization methods to vector fields or even tensor fields. 
In addition to changing the similarity measure, it is only necessary to adapt the spatial visualizations to analyze flow field ensembles. Suitable flow visualizations are already available in Voreen and can be easily exchanged due to the modular structure.

\section*{Acknowledgments}
This work was funded by the Deutsche Forschungsgemeinschaft (DFG, German Research Foundation) grant 260446826 (LI 1530/21-2). We would like to thank Verena Hörr, Simon Leistikow, Andreas Völker, and Raphael Wittkowski for their valuable ideas and feedback. 

\bibliographystyle{cag-num-names}
\bibliography{refs}

\end{document}